\documentclass[fleqn,usenatbib]{mnras}
\usepackage{newtxtext,newtxmath}

\usepackage[T1]{fontenc}
\usepackage{ae,aecompl}

\usepackage{graphicx}	
\usepackage{amsmath}	
\usepackage{array}
\usepackage{tabularx}
\usepackage{xcolor} 
\usepackage{multirow}
\usepackage{siunitx}

\newcommand{\Msun}{M_{\sun}}

\title[Deep searches for X-ray pulsations from Sco~X-1 and Cyg X-2]{Deep searches for X-ray pulsations from Scorpius X-1 and Cygnus X-2 in support of continuous gravitational wave searches}

\author[Galaudage et al.]{
Shanika Galaudage,$^{1,2}$\thanks{E-mail: shanika.galaudage@monash.edu }
Karl Wette,$^{3,2}$
Duncan K. Galloway,$^{1,2}$
and Chris Messenger$^{4}$
\\
$^{1}$School of Physics and Astronomy, Monash University, Clayton VIC 3800, Australia\\
$^{2}$ARC Centre of Excellence for Gravitational Wave Discovery (OzGrav), Hawthorn VIC 3122, Australia\\
$^{3}$Centre for Gravitational Astrophysics, Australian National University, Canberra ACT 2601, Australia\\
$^{4}$SUPA, School of Physics and Astronomy, University of Glasgow, Glasgow G12 8QQ, United Kingdom
}

\date{Accepted XXX. Received YYY; in original form ZZZ}

\pubyear{2021}

\begin{document}
\label{firstpage}
\pagerange{\pageref{firstpage}--\pageref{lastpage}}
\maketitle

\begin{abstract}
Neutron stars in low mass X-ray binaries are hypothesised to emit continuous gravitational waves that may be detectable by ground-based observatories. 
The torque balance model predicts that a higher accretion rate produces larger-amplitude gravitational waves, hence low mass X-ray binaries with high X-ray flux are promising targets for gravitational wave searches. 
The detection of X-ray pulsations would identify the spin frequency of these neutron stars, and thereby improve the sensitivity of continuous gravitational-wave searches 
by reducing the volume of the search parameter space. 
We perform a semi-coherent search for pulsations in the two low mass X-ray binaries Scorpius X-1 and Cygnus X-2 using X-ray data from the {\it Rossi X-ray Timing Explorer} Proportional Counter Array. 
We find no clear evidence for pulsations, and obtain upper limits (at 90\% confidence) on the fractional pulse amplitude, with the most stringent being 0.034\% for Scorpius X-1 and 0.23\% for Cygnus X-2.
These upper limits improve upon those of~\cite{Vaughan1994} by factors of $\sim 8.2$ and $\sim 1.6$ respectively.
\end{abstract}



\section{Introduction}\label{sec:intro}

Accreting neutron stars in low mass X-ray binaries (LMXBs) can be spun up due to the transfer of angular momentum via accretion. 
The measured spin frequencies of accreting neutron stars (typically in the range 100--700~Hz) are significantly below the expected neutron star break-up limit of $\sim \SI{1000}{Hz}$~\citep{Cook1994}. 
This observation suggests that there must be some mechanism by which the angular momentum is transferred away from these systems, as typical accretion rates over their long accretion lifetimes would otherwise be sufficient to achieve maximal spin rates. 
A possible mechanism~\citep{Bildsten1998,Chakrabarty2003} is the emission of gravitational waves from a rapidly-rotating neutron star that deviates from axisymmetry~\citep[e.g.][]{PapaPrin1978,Wagoner1984}. 

The frequency of the gravitational waves emitted by these sources is typically twice the spin frequency of the neutron star~\citep{VanDenBroeck2005}, and is modulated by the orbital motion of the star's binary companion.
The predicted strain is
\begin{equation}
    h_0 \approx \num{3e-27} \frac{R^{3/4}_{10}}{M^{1/4}_{1.4}}\bigg(\frac{F_\text{X}}{\SI{e8}{erg.cm^{-2}.s^{-1}}}\bigg)^{1/2}\bigg(\frac{\SI{1}{kHz}}{\nu}\bigg)^{1/2} \,,
    \label{eq:torquebalance}
\end{equation}
where $M$ is the mass of the neutron star ($M_{1.4}=M/1.4~\Msun$), $R$ is the radius of the neutron star ($R_{10}=R/\SI{10}{km}$), $F_\text{X}$ is the observed X-ray flux, and $\nu$ is the spin frequency of the neutron star~\citep{Watts2008}. 
These gravitational waves are expected to be emitted persistently~\citep[e.g.][]{Prix2009:GrvWvSpnNtSt} and are referred to as continuous gravitational waves (CWs).
They are a potential target for current-generation gravitational wave observatories such as LIGO~\citep{AdLIGO} and Virgo~\citep{Virgo}, as well as for future ground-based detectors.
Numerous searches for continuous gravitational waves from LMXBs have been performed to date~\citep{LIGO2007:SrcPrGrvWUIsSScXRSLSR,LIGO2007:UpLmMBckGrvWv,LIGOVirg2015:DrcSrGrvWScXInLD,LIGOVirg2017:SrGrvWScXFAdLObsRHMM,LIGOVirg2017:DrcLPrGrvWAdLFObR,LIGOVirg2017:ULGrvWScXMdCrsSAdLD,MeadEtAl2017:SrCntGrvWScXXJ1LSSR,LIGOVirg2019:SGrvWScXSAdLObsRImHMM,LIGOVirg2019:DrcLPrGrvWUDAdLFTObR,MiddEtAl2020:SGrvWFLMXBnSAdLObsRImHMM}.

To search for continuous gravitational waves, we require a template describing the expected signal waveform.
To construct such a template, we need information about the neutron star, such as its position in the sky and frequency evolution as a function of time. 
For neutron stars in binary systems, we additionally need to correct for the orbital motion of the system. 
Uncertainties in the orbital parameters result in a multiplicity of  templates which must be searched to recover the true signal waveform. 
The greater the uncertainty in the binary system parameters, the larger the number of templates needed to search the parameter space at a given template resolution. 
For sufficiently large parameter spaces, the significant computational cost of the search requires the use of sub-optimal semi-coherent search strategies which sacrifice sensitivity for reduced computational cost~\citep[e.g.][]{Messenger2011,Leaci2015}.
Conversely, sufficiently precise constraints on the orbital parameters of the binary and spin frequency of the neutron star would reduce the search parameter space volume to a negligible number of templates and allow for an optimal fully-coherent search, thereby improving the sensitivity to a continuous wave signal.

The LMXBs Scorpius X-1 (Sco~X-1) and Cygnus X-2 (Cyg~X-2) are considered to be promising candidates for the detection of continuous gravitational waves.
They have high X-ray fluxes ($\gtrsim \SI{e9}{erg.cm^{-2}.s^{-1}}$) and precise constraints on the orbital period, radial velocities and epoch of inferior conjunction~\citep{Wang2018,Premachandra2016} which reduces the search parameter space. 
The spin frequency of these sources are unknown, however~\citep{Damle1988,Wood1991,Vaughan1994,Manchanda2005}.
A measurement of the spin frequency would greatly reduce the parameter space volume of continuous gravitational wave searches targeting these neutron stars. 
A detection of X-ray pulsations would immediately provide such a measurement.

Searches for X-ray pulsations face similar challenges to continuous gravitational wave searches. These include accounting for the Doppler modulation of the signal phase due to the binary orbit; as a result, the computational cost of the search increases much more rapidly with observation time than does the gain in sensitivity~\citep{Messenger2011,Leaci2015}, making a fully-coherent search computationally costly if not entirely infeasible.
In addition, the search must consider the unknown variations in the spin frequency driven by the varying accretion rate, known as spin wandering~\citep[e.g.][]{Mukherjee2018}.
The timescale of the frequency variation due to spin wandering is highly uncertain, and limits the time-span of a fully-coherent search for X-ray pulsations, and hence its achievable sensitivity.

In this paper we perform searches of X-ray data from Sco~X-1 and Cyg~X-2 using a semi-coherent method developed by~\cite{Messenger2011}, and previously used to search for X-ray pulsations from LMXBs in~\cite{Messenger_Patruno2015} and~\cite{Patruno2018}.  In Section~\ref{sec:xray_obs} we outline the X-ray data selection and processing methods. In Section~\ref{sec:semicoherent} we discuss the details of the search method and the requirements of the search. In Section~\ref{sec:results} we report the findings of our searches. Finally in Section~\ref{sec:discussion} we discuss the implications of our results and possible avenues for future work.
 
\section{X-ray observations}\label{sec:xray_obs}

The X-ray data used for Sco~X-1 and Cyg~X-2 were collected with the Proportional Counter Array (PCA) onboard the {\it Rossi X-ray Timing Explorer}\/ ({\it RXTE};~\citealt{Jahoda2006}). 
The PCA consists of five identical proportional counter units (PCUs), sensitive to X-ray photons in the energy range 2--60~keV, and with a total effective area of $\sim \SI{7000}{cm^2}$. The instrument collects data in two ``standard'' modes, as well as additional user-defined data modes\footnote{\href{url}{https://heasarc.gsfc.nasa.gov/docs/xte/abc/modes\_sorted.html}} offering different combinations of time and energy resolution.
The volume of data available for both X-ray sources is very large; given that we are limited by computational cost, we select and analyse a subset of the total data available (see Section~\ref{sec:DataSelection}). 

\subsection{Observation data}\label{sec:obs}

The {\it RXTE}/PCA data were collected using a variety of datamodes. For X-ray pulsation searches, we require data modes suitable for measuring variations in the light curve at high resolution ($< \SI{1}{ms}$). There are three key modes we used to obtain the light curves for Sco~X-1 and Cyg~X-2: ``EVENT'' modes, ``BINNED'' modes and ``SINGLE BIT'' modes. 
The observation IDs of the data used for this analysis are listed in Appendix~\ref{sec:dataset-IDs}. We used $\sim \SI{143}{hr}$ of data spanning from January 2, 1998 to Dec 26, 2010.

Since Sco~X-1 is an extremely bright source ($> \SI{e5}{counts.s^{-1}}$), modes which exceed the telemetry limit ($\SI{40}{kbit.s^{-1}}$) result in gaps in the data. These modes are excluded from this analysis.
For some observations, a unique approach was used to collect and process some of the data, involving using multiple event analyzers to collect information about events detected by different PCUs. Additionally, as reported by~\cite{jones08}, 
there is a significant probability of multiple detections in separate regions of the detectors, which would normally be excluded by the anti-coincidence electronics. To recover a complete time-series for Sco~X-1 for the affected observations, we summed the time-series for both datamodes, as well as two counts for each of the coincident (``2-LLD'') events.
This procedure required the use of 
DSTOOLS,\footnote{E.~Morgan, private communication.} 
a set of data processing tools which
process compressed ``DS''-format files extracted from packet data obtained independently of the FITS data provided by the Guest Observer Facility.

For Cyg~X-2, which is several orders of magnitude fainter than Sco~X-1, we instead used the FITS data and created lightcurves using FTOOLS~\citep{FTOOLS}.

\subsection{Data corrections}\label{sec:data}

There are a number of factors required to correct the X-ray data to obtain the best estimate of the light curve. These factors include correcting for the motion of the Earth in the Solar System, the instrumental effects of {\it RXTE}'s PCA, and the deadtime. 
We used the JPL DE200 ephemeris to apply the barycentric corrections~\citep{Standish1990}. The instrumental effects of the PCA need to be accounted for in order to determine the count rate. These include: the offset angle from the source of the instrument, the number of proportional counter units (PCUs) in operation, and the time where the detctor is not processing an event due to processing another referred to as the dead-time. These corrections are given by
\begin{equation}
\mu = \lambda \left( \frac{1}{1 - \theta/\SI{1}{deg}} \right) \left( \frac{1}{1 - \tau_{\text{dtf}}} \right) \left( \frac{1}{n_{\text{PCU}}} \right) \,,
\end{equation}
where $\mu$ is the best estimate of the count rate, $\lambda$ is the detected count rate, $\theta$ is the offset angle in degrees, $\tau_{\text{dtf}}$ is the dead-time fraction, and $n_{\text{PCU}}$ is the number of PCUs in operation. 
Note that the photon detection efficiency is inversely proportional to offset angle; see Fig. 32 in~\cite{Jahoda2006}.
The dead-time is proportional to the photon count. For Sco X-1 and Cyg X-2 the typical dead-time fractions are $\sim 0.3$ and $\sim 0.03$ respectively.

\subsection{Data selection}\label{sec:DataSelection}

\begin{figure*}
    \centering
    \includegraphics[width=\columnwidth]{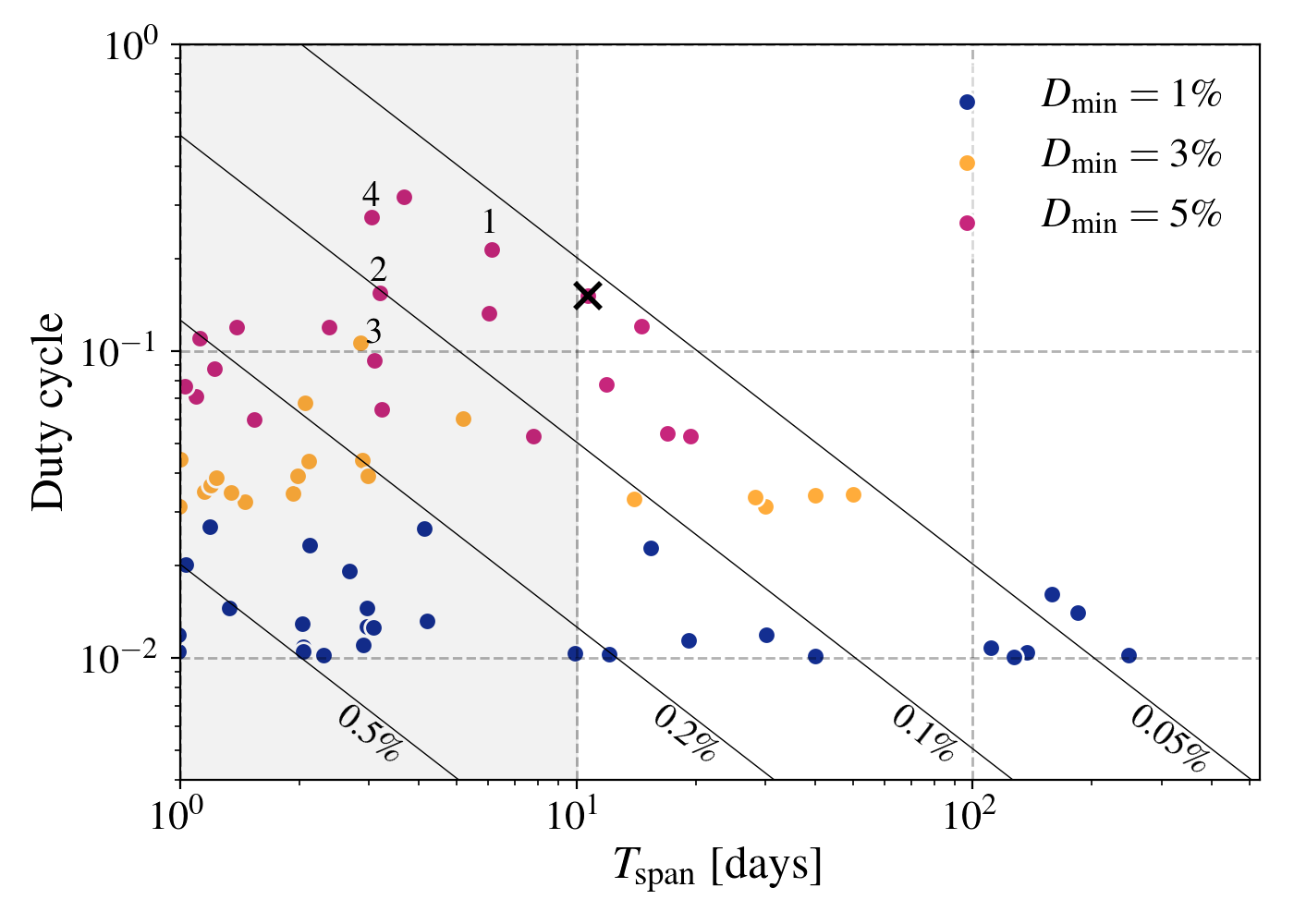}
    \includegraphics[width=\columnwidth]{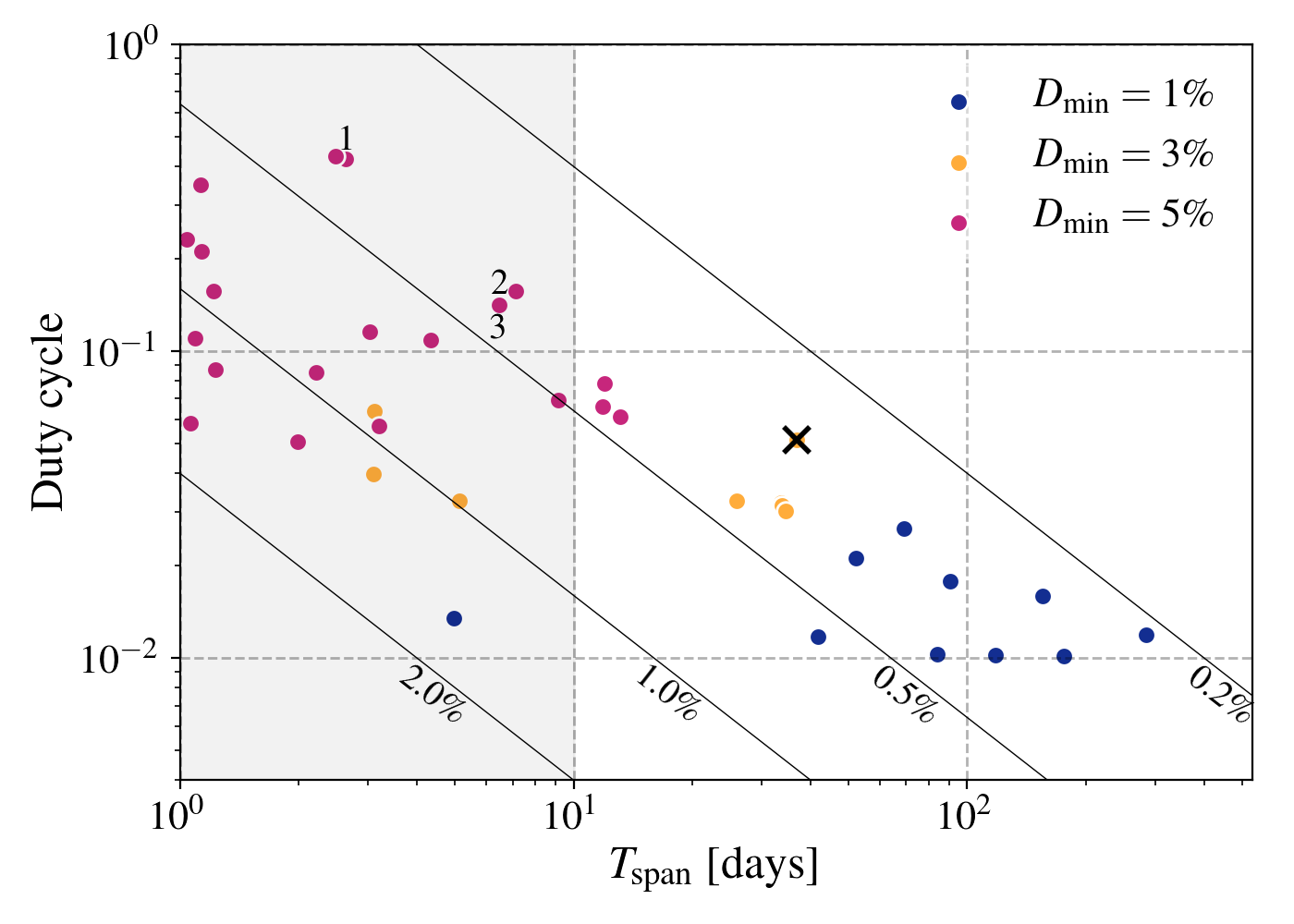}
    \caption{Duty cycle plotted against total time-span for selected subsets of {\it RXTE}/PCA data for Sco~X-1 (left) and Cyg~X-2 (right). 
    The dots represent datasets with a set time-span and duty cycle, and the different colours represent the set duty cycle cut-off for that dataset (navy: 1\%; yellow: 3\%; and magenta: 5\%). 
    The black contour lines define the fractional detectable pulse amplitude.
    The numbers correspond to the datasets selected for searches limited to a 10 day coherence time. Properties of these datasets are listed in Tables~\ref{tab:scox1results} and \ref{tab:cygx2results}.
    The black cross indicates the dataset selected for the search that is not limited by the coherence time.
    The grey shaded region corresponds to a time-span of $\leq \SI{10}{days}$.}
    \label{fig:dutycycle}
\end{figure*}

For each LMXB we performed two complementary searches for X-ray pulsations, based on different assumptions about the spin evolution. 
For the first search, we assume that the spin frequency of the neutron stars have an unknown variation due to spin wandering; this constraint restricts the span of our datasets to the maximum spin-wandering timescale of the source. 
For the second search, we assume that spin wandering is absent (or has no effect on the spin frequency), and hence the spin frequency is constant.
The sensitivity achieved with the latter search is constrained only by the amount of data available and the computational time.

Within a time-span $T_{\text{span}}$, we may find $M$ X-ray observations of average duration $T$.
The total on-source time $n T$ is limited by $T_{\text{span}}$; we call the ratio $M T / T_{\text{span}}$ the duty cycle $D$.
(For simplicity, we assume that there are no data gaps \emph{within} each X-ray observation.)
This duty cycle is typically quite small; constraints include occultation of the X-ray source by the Earth during {\it RXTE}'s short ($\sim \SI{100}{min}$) low-Earth orbit, times when the satellite passed through the South Atlantic Anomaly, and times when an X-ray source cannot be observed due to its proximity to the Sun.
In addition, observations of X-ray sources are also generally scheduled in response to one or more observing or monitoring proposals, and not necessarily with the goal of maintaining a high duty cycle.

The relationship between the time-span $T_{\text{span}}$, X-ray segment duration $T$, duty cycle $D$, detectable fractional pulse amplitude $A$, and photon count rate $\mu$ is given by
\begin{equation}
    A \sim 8\mu^{-1/2} (T_{\text{span}} T D)^{-1/4} \,.
    \label{eq:simpleA}
\end{equation}
This expression is a simplified version of that given in~\cite{Messenger_Patruno2015}. Using this estimate as a guide, we calculate the detectable fractional pulse amplitude obtained for a dataset of a given time-span and duty cycle. 
We partition the data with a number of different cut-offs for duty cycle; for example, setting a cut-off of 1\% for duty cycle would mean that each segment would have a duty cycle of at least 1\%.  
Figure~\ref{fig:dutycycle} shows duty cycle plotted against total time-span for selected subsets of {\it RXTE}/PCA data for Sco~X-1 and Cyg~X-2 respectively. 
Using these estimates, we can identify the best datasets to analyse to obtain the most sensitive search possible from the available data.\footnote{Due to our simplifying assumption that there are no data gaps within each X-ray observation, the ultimate sensitivity realised by the searches may be lower by a factor of 2-4 than what is given in Fig.~\ref{fig:dutycycle}.
Nevertheless, the computational cost of the search is much more sensitive to $D$ -- the duty cycle of segments within a given time-span $T_{\text{span}}$ -- than the presence of data gaps within a segment of duration $T \ll T_{\text{span}}$.
Hence the assumption of no data gaps should not noticeably affect the ranking of datasets in order to achieve the best sensitivity.
}

For the first search, where we consider the effects of spin wandering, the duration of the datasets used for the searches is limited to the calculated spin wandering timescales. 
\citet{Mukherjee2018} estimated the range of spin wandering timescales for Sco~X-1 to be 5--80~days. 
Using an estimate for the accretion rate based on the luminosity, and neutron star mass and radius, we assume that the spin wandering timescales will be similar to that of Cyg X-2.
We select a conservative value of $10$ days for Sco~X-1 and Cyg~X-2. This duration is the maximum over which the variation in spin frequency can be well modelled.
For the second search (assuming no spin wandering), we select the dataset that maximises the duty cycle whilst minimising the computational cost. 

While Figs.~\ref{fig:dutycycle} indicates the most sensitive datasets, they do not account for the computational cost of analysing those datasets, which may be infeasible for large $T_\text{span}$.
We therefore need to make pragmatic choices of datasets which achieve almost the best sensitivity, but at significantly reduced computational cost.
For example, in Fig.~\ref{fig:dutycycle} we select datasets that does not necessarily give the smallest percentage fractional pulse amplitude $A$, but instead has a shorter $T_\text{span}$ with a comparable value of $A$. 
This selection is made to reduce the computational cost: the difference is sensitivity between these two datasets is only a factor of $\approx 1.07 $, whereas the computational cost differs by a factor $> 10$. 
Note that the segment length $T$ can also vary for a given dataset, and for datasets with higher duty cycles we generally have longer segments of data; from Eq.~\ref{eq:simpleA} we can see that we can therefore improve sensitivity by increasing $T$.
Once datasets are selected, robust estimates of the sensitivity of each of the searches is determined using a variant of the analytic calculation given in~\cite{Wette2012}.

\section{Search method}\label{sec:semicoherent}

We used a semi-coherent search method to search for X-ray pulsations from Sco~X-1 and Cyg~X-2.
Briefly, the method partitions the X-ray data into segments of length $T$, performs a fully-coherent analysis of each segment, then combines results from each segment such that the frequency evolution of the signal -- though not necessarily its phase evolution -- is self-consistent with a single set of parameters over the total time-span $T_{\text{span}}$ of the search; this yields a detection statistic $\Sigma$.
A detailed description of the method can be found in~\cite{Messenger2011} and~\cite{Messenger_Patruno2015}.
We briefly summarise parts of the search method in this section.

\subsection{Signal model}

X-ray pulsations from an LMXB system are modelled by a time series
\begin{equation}
    {r}_j({\theta}) = R\big\{1+A\sin[\phi _j(\mathbf{\theta}) + \beta]\big\} \,,
    \label{eq:signal}
\end{equation}
where $j$ indexes time, ${r}_j({\theta})$ is the detected counts in the $j$th time bin, $R$ is the expected background counts per time bin, $A$ is the pulsed fraction of our signal, $\phi _j ({\theta})$ is the signal phase, and $\beta$ is a reference rotation phase of the signal. 
The Doppler modulation of phase $\phi _j({\theta})$ is given by
\begin{equation}
    \phi _j({\theta}) = 2\pi\nu\big\{t_j - t_0 - a \sin[\Omega(t_j-t_0)+\gamma]\big\},
    \label{eq:phase} \,
\end{equation}
where $\nu$ is the spin frequency, $t_j$ is the time of the $j$th time bin, $t_0$ is a reference time, $a$ is the projected semi-major axis of the orbit, $\Omega=2\pi/P_{\text{orb}}$ is the orbital frequency, $P_{\text{orb}}$ is the orbital period, $\gamma=\Omega(t_0-T_{\text{asc}})$ is the orbital phase, and $T_{\text{asc}}$ is the time of ascension.

\subsection{Parameter space}

\begin{table}
\caption{Search parameter ranges for Sco~X-1.}
\label{tab:parameter_scox1}
\begin{tabularx}{\linewidth}{Xcccc} 
	\hline
	Parameter & Dataset & Min & Max & Units\\
	\hline
	$\nu$    & all & 100 & 700   & Hz \\
	$P_{\text{orb}}$   & all & 68023 & 68024 & s    \\
	$a$  & all & 1.45 & 3.25 & lt-s    \\
	\multirow{5}{*}{$T_{\text{asc}}$}  & 1 & 568041816 & 568042347 & \multirow{5}{*}{GPS s}    \\
	& 2 & 580694264 & 580694775 & \\
	& 3 & 583551268 & 583551775 & \\
	& 4 & 568177864 & 568178394 & \\
	& no spin & 600081075 &  600081565 & \\
	\hline		
\end{tabularx}
\end{table}

\begin{table}
\caption{Search parameter ranges for Cyg~X-2.}
\label{tab:parameter_cygx2}
\begin{tabularx}{\linewidth}{Xcccc} 
	\hline
	Parameter & Dataset & Min & Max & Units\\
	\hline
	$\nu$  & all  & 100 & 700   & Hz \\
	$P_{\text{orb}}$  & all & 850575 & 850601 & s    \\
	$a$  & all & 12.47 & 14.04 & lt-s    \\
	\multirow{4}{*}{$T_{\text{asc}}$}  & 1 & 551418623 & 551421914 & \multirow{4}{*}{GPS s}    \\
	& 2 & 974156786 & 974168010 & \\
	& 3 & 976708513 & 976719809 & \\
	& no spin & 974156786 & 974168010 & \\
	\hline
\end{tabularx}
\end{table}

The search builds a parameter space of template waveforms to search in order to correct for the Doppler modulation due to the orbital motion of the LMXB system [Eq.~\eqref{eq:signal}], which is parameterised by $\nu$, $a$, $\Omega$, and $\gamma$.
For example, the amplitude of the Doppler modulation is proportional to $\nu a$, and its period is equal to $P_\text{orb}$.
The volume of the parameter space is defined by the range of the orbital parameters that must be searched over; the larger the uncertainty in a parameter, the larger the volume of the parameter space.

Tables~\ref{tab:parameter_scox1}~and~\ref{tab:parameter_cygx2} list the parameter ranges for Sco~X-1~\citep{Wang2018} and Cyg~X-2~\citep{Premachandra2016} respectively.
Given that the spin frequency of the target sources is unknown, we chose the range for this parameter to be 100--700~Hz. 
This range follows from the assumption that the spin frequencies of these sources will lie within the current known distribution of spin frequencies for most accreting neutron star systems~\citep{Patruno2017}.
For all datasets the search range of the spin frequency, orbital period and projected semi-major axis remain the same. 
The time of ascension and its uncertainty are propagated to the epochs of the observed Sco~X-1 and Cyg~X-2 data; hence a value of the time of ascension is given for each dataset. 
The range of the time of ascension also varies for each dataset since its uncertainty grows with every orbit from the measured value.

The search method chooses template parameters $(f, a, \Omega, \gamma)$ at random~\citep{MessEtAl2009:RnTmpBRlLtCvr} within the ranges given in Tables~\ref{tab:parameter_scox1}~and~\ref{tab:parameter_cygx2}.
The templates are associated with a maximum mismatch $m_\text{max}$; this is the maximum loss in signal-to-noise ratio that will be permitted, due to the fact that no random template will ever exactly match the parameters of the signal.
Because the templates are chosen at random, the condition that the signal will be close enough to a random template to satisfy the maximum mismatch condition is guaranteed only for a fraction $\eta < 1$ of the parameter space.
There is therefore a non-zero probability that the random templates do not cover some non-empty subset of the parameter space.

The number of semi-coherent templates required to obtain a fraction $\eta$ of parameter-space coverage is given by
\begin{align}
\label{eq:templates}
\begin{split}
n = & \log \bigg(\frac{1}{1-\eta}\bigg) \frac{\pi^4 T^4 T_{\text{span}}}{25920 m_{\text{max}}^2}(\nu^4_{\text{max}} - \nu^4_{\text{min}})(a^3_{\text{max}} - a^3_{\text{min}})\\
&\quad \times~(\Omega^4_{\text{max}} - \Omega^4_{\text{min}})(\gamma _{\text{max}} - \gamma _{\text{min}}) \,.
\end{split}
\end{align}
We can approximate the number of templates for Sco~X-1 as
\begin{equation}
n \approx \num{6.2e8} \bigg(\frac{T}{\SI{512}{sec}}\bigg)^4\bigg(\frac{T_{\text{span}}}{\SI{10}{days}}\bigg) \,,
\label{eq:templates1}
\end{equation}
and for Cyg~X-2 as
\begin{equation}
n \approx \num{5.1e6} \bigg(\frac{T}{\SI{512}{sec}}\bigg)^4\bigg(\frac{T_{\text{span}}}{\SI{10}{days}}\bigg) \,,
\label{eq:templates2}
\end{equation}
for a typical observation span where the mismatch $m = 0.01$, and the coverage $\eta = 0.9$. The actual number of templates used for each dataset is given in Tables~\ref{tab:scox1results}~and~\ref{tab:cygx2results}.

\section{X-ray Search Results}\label{sec:results}

We searched the archival {\it RXTE}/PCA data, prepared in Section~\ref{sec:xray_obs}, for X-ray pulsations from Sco~X-1 and Cyg~X-2 using the search method detailed in Section~\ref{sec:semicoherent}.
No clear evidence for pulsations was found. 

We present the following sets of results: searches that assume spin wandering, and hence use datasets limited to the spin wandering timescale; and searches which assume no spin wandering.
Details about the properties of each dataset are provided in Tables~\ref{tab:scox1results},~\ref{tab:cygx2results} and~\ref{tab:nospin_results}. These properties include: the GPS start time of each dataset $t$, the total on-source time within the dataset $T_{\text{obs}}$, the time-span of the dataset $T_{\text{span}}$, the total number of photons $\mathcal{N}$, the time-span of a single segment $T$, the number of segments $M$, the number of semi-coherent templates $n$, the upper limit of fractional pulse amplitude at 1\% false alarm probability and 10\% false dismissal probability (90\% confidence) $A^{1\%}_{10\%}$, the 1\% false alarm threshold on the detection statistic $\Sigma^{1\%}$, and the maximum value of the detection statistic found by the search $\Sigma^{*}$.

\subsection{Assuming spin wandering}

\begin{table*}
\caption{Data parameters, estimated sensitivities, and upper limits for Sco~X-1. 
Columns are: dataset number, GPS start time of dataset, total on-source time within the dataset, time-span of dataset, number of photons, time-span of a single segment, number of segments, number of semi-coherent templates, upper limit of fractional pulse amplitude at 1\% false alarm probability and 10\% false dismissal probability (90\% confidence), 1\% false alarm threshold on the detection statistic, largest value of detection statistic found by the search.}
\label{tab:scox1results}
\begin{tabularx}{\linewidth}{l@{\extracolsep{\fill}}cccccccccc}
    \hline
    No. & $t$ [GPS~s] & $T_{\text{obs}} / 10^5$ [s] & $T_{\text{span}} / 10^5$ [s] & $\mathcal{N} / 10^{9}$ & $T$ [s] & $M$ & $n / 10^{9}$  & $A^{1\%}_{10\%}$ & $\Sigma^{1\%}$ & $\Sigma ^{*}$ \\
    \hline
    1 & 567787966 & 0.89 & 5.45 & 3.17 & 528 & 153 &  4.95 & 0.035\% & 510.15 & 486.06  \\
    2 & 580523651 & 0.46 & 2.82 & 1.50 & 697 &  55 & 10.63 & 0.036\% & 248.78 & 230.08  \\
    3 & 583396353 & 0.30 & 2.73 & 1.01 & 579 &  46 &  6.24 & 0.041\% & 220.22 & 197.31  \\
    4 & 613041597 & 0.61 & 2.65 & 1.45 & 699 &  77 & 10.99 & 0.041\% & 311.79 & 297.13  \\
    \hline
\end{tabularx}
\end{table*}

\begin{table*}
\caption{Data parameters, estimated sensitivities, and upper limits for Cyg~X-2. Columns are as for Table~\ref{tab:scox1results}.}
\label{tab:cygx2results}
\begin{tabularx}{\linewidth}{l@{\extracolsep{\fill}}cccccccccc}
    \hline
    No. & $t$ [GPS~s] & $T_{\text{obs}} / 10^5$ [s] & $T_{\text{span}} / 10^5$ [s] & $\mathcal{N} / 10^{6}$ & $T$ [s] & $M$ & $n / 10^{9}$  & $A^{1\%}_{10\%}$ & $\Sigma^{1\%}$ & $\Sigma ^{*}$ \\
    \hline
    1 & 551735827 & 0.60 & 2.33 & 3.64 & 2402 & 30 & 38.28 & 0.61\% & 175.80 & 155.43  \\
    2 & 974186811 & 0.78 & 5.65 & 21.39 & 2111 & 39 & 52.79 & 0.27\% & 205.64 & 195.56 \\
    3 & 976792264 & 0.73 & 5.91 & 24.57 & 3288 & 40 & 39.76 & 0.26\% & 207.87 & 207.56  \\
    \hline
\end{tabularx}
\end{table*}

\begin{table*}
\caption{Data parameters, estimated sensitivities, and upper limits assuming no spin wandering. Columns are as for Table~\ref{tab:scox1results}, except that column 1 lists the source name.}
\label{tab:nospin_results}
\begin{tabularx}{\linewidth}{l@{\extracolsep{\fill}}cccccccccc}
    \hline
    Source & $t$ [GPS~s] & $T_{\text{obs}} / 10^5$ [s] & $T_{\text{span}} / 10^5$ [s] & $\mathcal{N} / 10^{9}$ & $T$ [s] & $M$ & $n / 10^{9}$  & $A^{1\%}_{10\%}$ & $\Sigma^{1\%}$ & $\Sigma ^{*}$ \\
    \hline
    Sco~X-1 & 599635471 & 1.50 &  9.27 & 4.25 & 707  & 199 & 16.71 & 0.034\% & 632.59 & 594.18  \\
    Cyg~X-2 & 974186811 & 1.51 & 31.97 & 0.05 & 3499 & 79 & 46.97 & 0.23\%  & 322.97 & 290.40 \\
    \hline
\end{tabularx}
\end{table*}

\begin{figure}
	\includegraphics[width=\columnwidth]{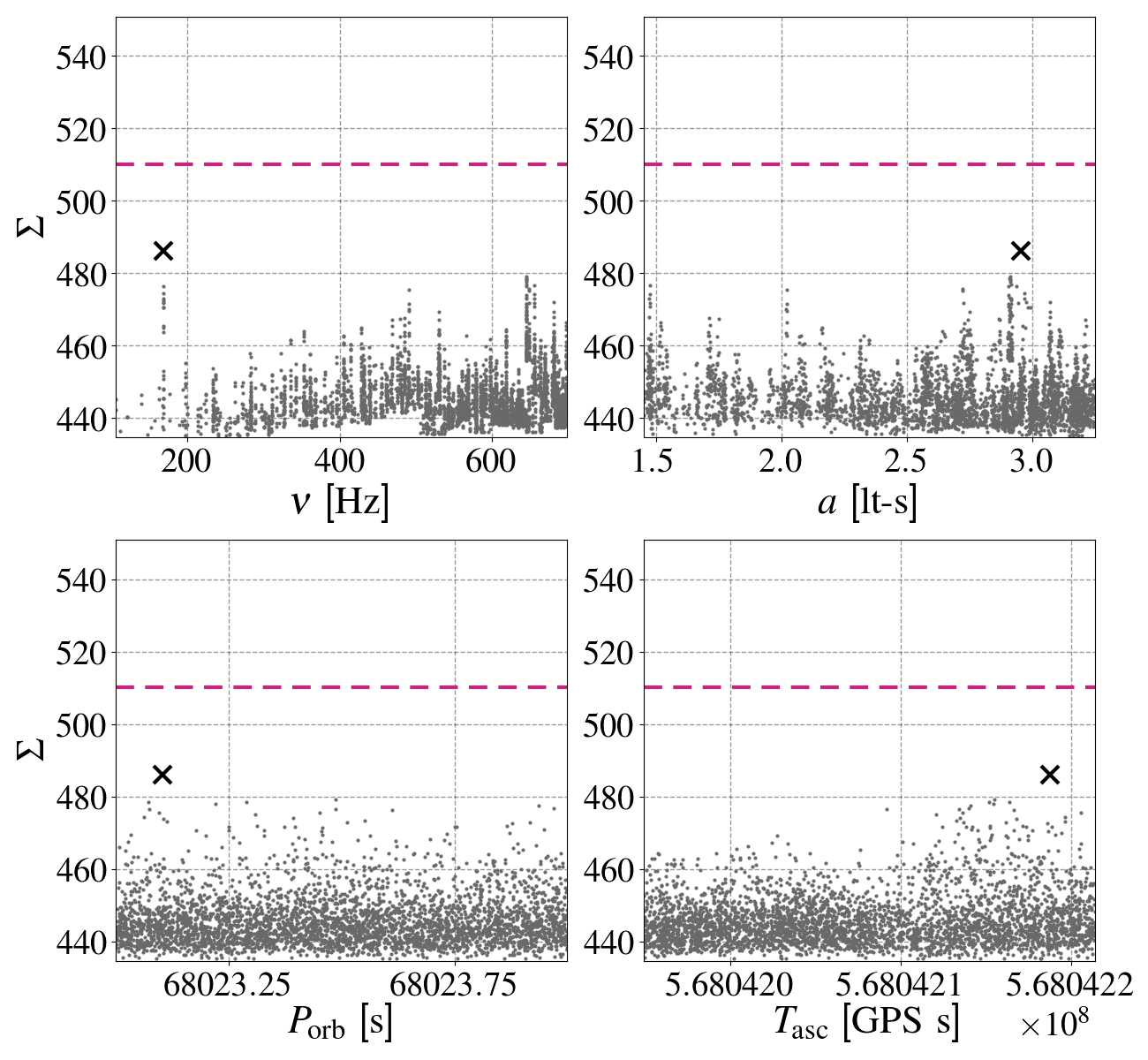}
    \caption{Detection statistic $\Sigma$ plotted against the search range of frequency $\nu$, projected semi-major axis $a$, period $P_{\text{obs}}$ and time of ascension $T_{\text{asc}}$ for dataset~1 of Sco~X-1; see Table~\ref{tab:scox1results}. The black cross indicates the template with the largest detection statistic found by the search. The dashed horizontal line at $\Sigma ^{1\%} = 510.15$ indicates the 1\% false alarm threshold on the detection statistic and corresponds to a fractional pulse amplitude of 0.035\%. To parallelise the search, the parameter space is divided into partitions; the figure plots the 100 largest detection statistics recorded in each partition.
    }
  \label{fig:dataset1_scox1}
\end{figure}

\begin{figure}
	\includegraphics[width=\columnwidth]{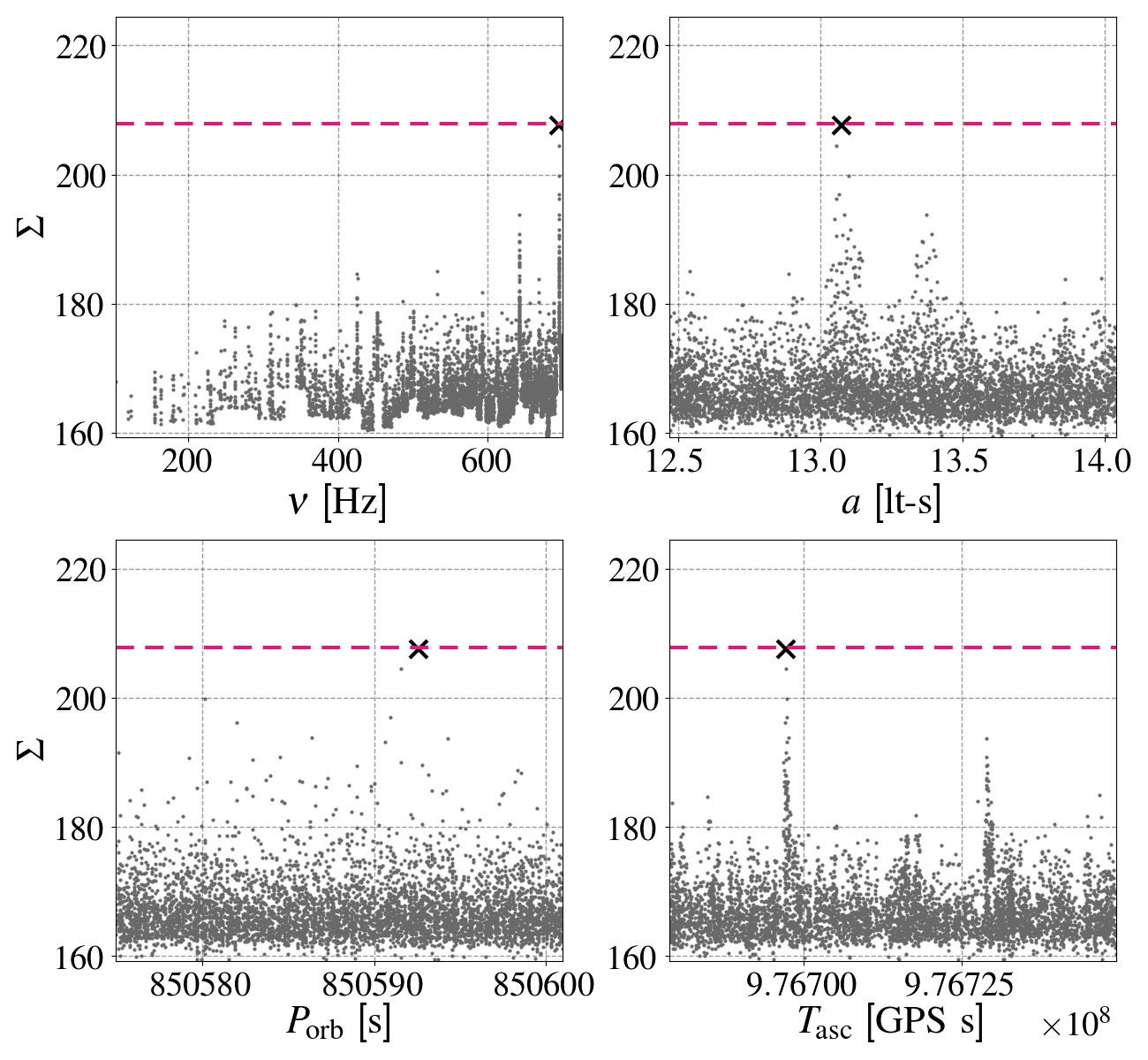}
    \caption{Same as Fig.~\ref{fig:dataset1_scox1}, but for dataset~3 of Cyg~X-2; see Table~\ref{tab:cygx2results}. The dashed horizontal line at $\Sigma ^{1\%} = 207.87$ indicates the 1\% false alarm threshold on the detection statistic and corresponds to a fractional pulse amplitude of 0.26\%. 
    }
  \label{fig:dataset3_cygx2}
\end{figure}

\begin{figure}
	\includegraphics[width=\columnwidth]{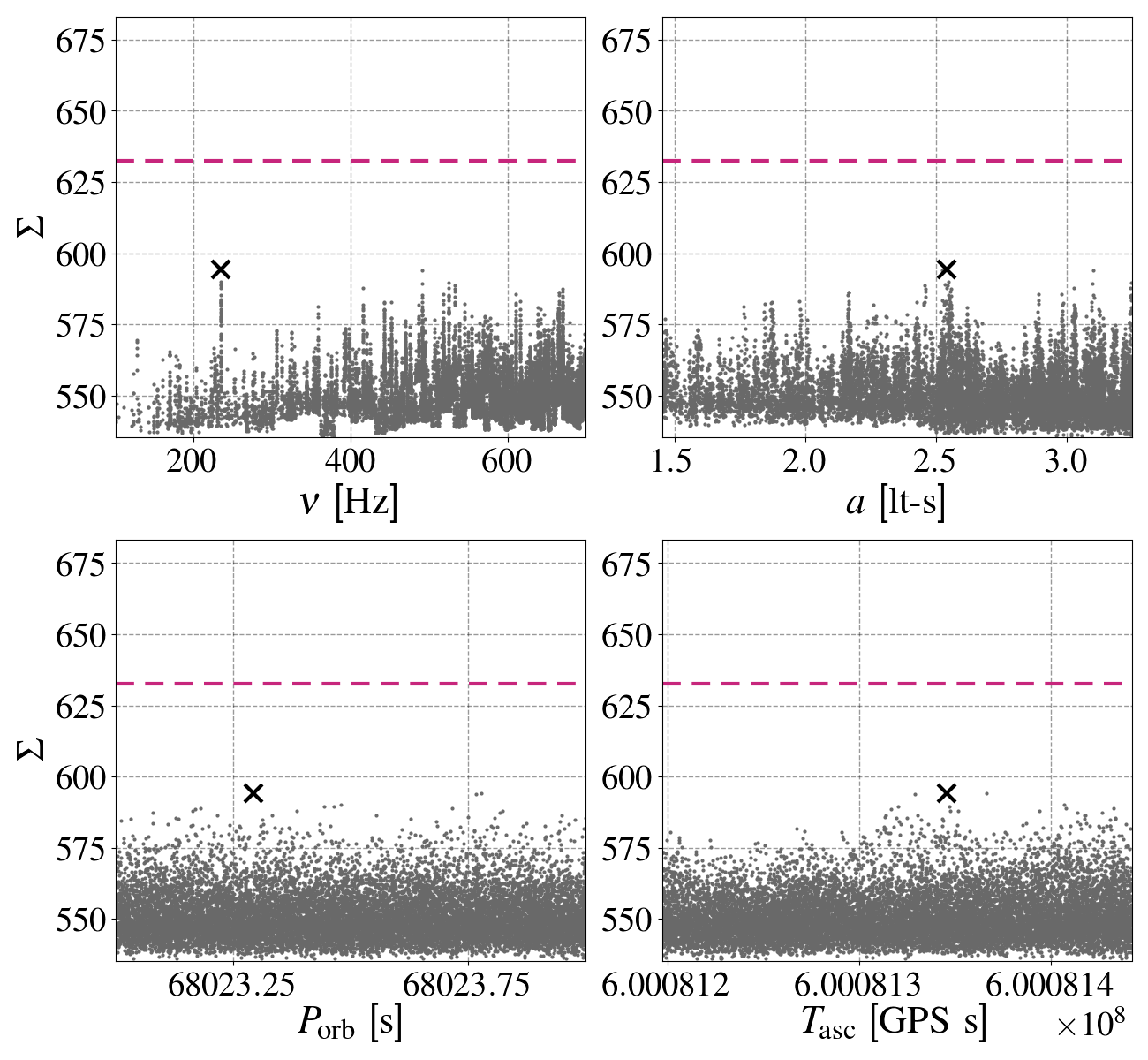}
    \caption{Detection statistic $\Sigma$ plotted against the search range of frequency $\nu$, projected semi-major axis $a$, period $P_{\text{obs}}$ and time of ascension $T_{\text{asc}}$ for a search for Sco~X-1 assuming no spin wandering; see Table~\ref{tab:nospin_results}. The black cross indicates the template with the largest detection statistic found by the search. The dashed horizontal line at $\Sigma ^{1\%} = 632.59$ indicates the 1\% false alarm threshold on the detection statistic and corresponds to a fractional pulse amplitude of 0.034\%. The 100 largest detection statistics recorded in each parameter space partition are plotted.}
    \label{fig:scox1_no_spin}
\end{figure}

\begin{figure}
	\includegraphics[width=\columnwidth]{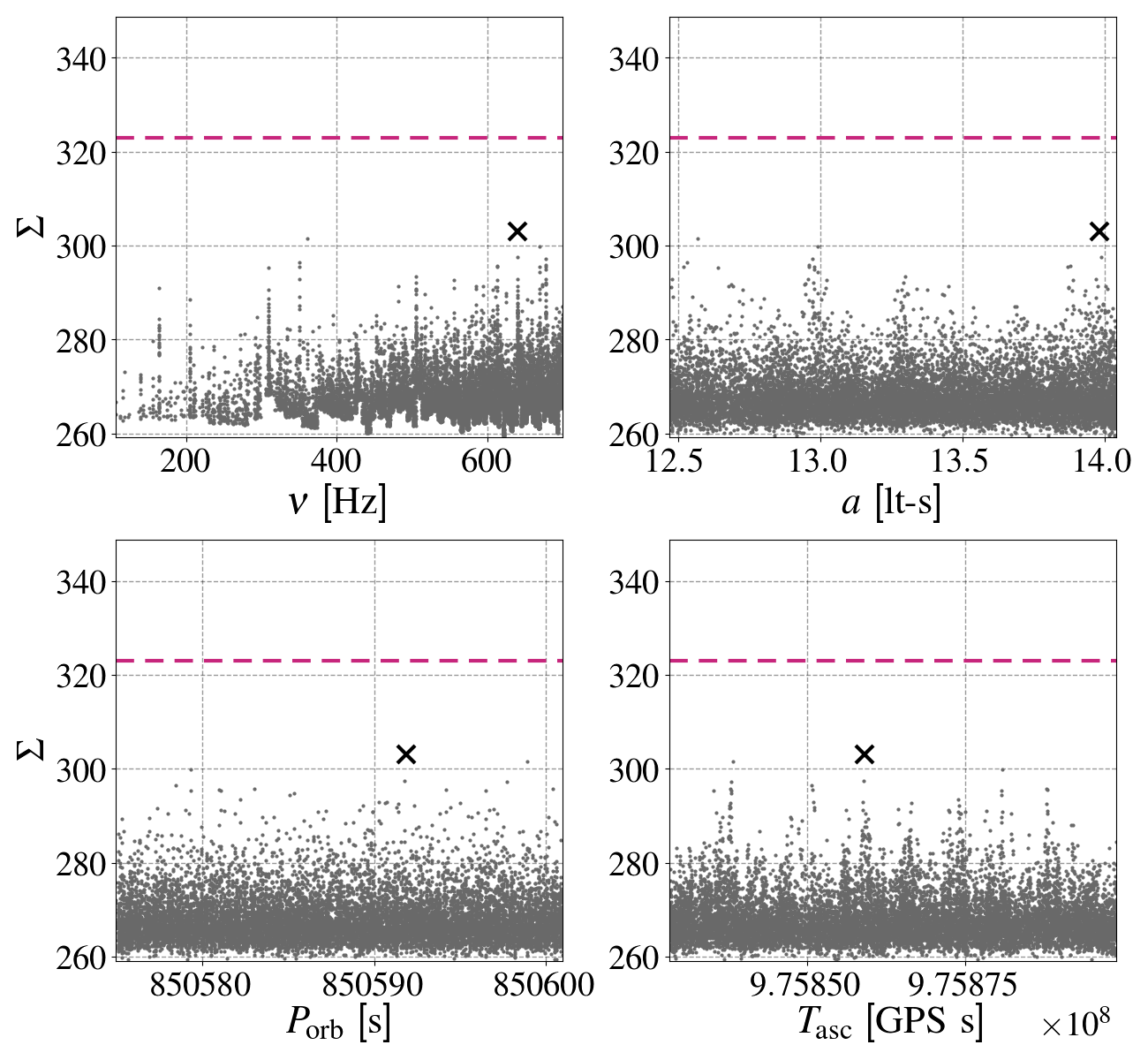}
    \caption{Same as Fig.~\ref{fig:scox1_no_spin} but for Cyg~X-2. The dashed horizontal line at $\Sigma ^{1\%} = 322.97$ indicates the 1\% false alarm threshold on upper limit and corresponds to a fractional pulse amplitude of 0.23\%.}
    \label{fig:cygx2_no_spin}
\end{figure}

The properties of the 4 datasets searched for Sco~X-1, assuming spin wandering, are given in Table~\ref{tab:scox1results}. 
We set upper limits on the detectable fractional pulse amplitude of each dataset; the most stringent result for Sco~X-1 is a fractional pulse amplitude of 0.035\% for dataset~1. For this dataset, in Fig.~\ref{fig:dataset1_scox1} we plot the largest values of the detection statistic $\Sigma$ found by the search against the four parameters ($\nu$, $a$, $P_\text{orb}$ and $T_\text{asc}$) used to define the parameter space.

The properties of the 3 datasets searched for Cyg~X-2, again assuming spin wandering, are given in Table~\ref{tab:cygx2results}. 
The most stringent upper limit for Cyg~X-2 is a fractional pulse amplitude of 0.26\% for dataset 3; for this dataset, we plot in Fig.~\ref{fig:dataset3_cygx2} the largest values of $\Sigma$ against $\nu$, $a$, $P_\text{orb}$ and $T_\text{asc}$.
The largest candidate in this dataset has $\Sigma = 207.56$, just below the 1\% false alarm threshold $\Sigma ^{1\%} = 207.87$; its parameters are $\nu = \SI{695.09}{Hz}$, $a = \SI{13.08}{lt{-}s}$, $P_{\text{orb}} = \SI{850592.58}{s}$, and $T_{\text{asc}} = \SI{976697337.62}{GPS~s}$.

A follow-up analysis for this sub-threshold candidate following the approach of~\cite{Patruno2018}, where the segment time-span $T$ is increased to improve sensitivity, is unfortunately not possible in this case.
The Cyg X-2 data was not taken at 100\% duty cycle, but in a series of discontinuous observations (see Appendix~\ref{sec:dataset-IDs}).
The initial search of dataset 3 used $T = \SI{3288}{s}$ (see Table~\ref{tab:cygx2results}), which is approximately the length of the longest observation of Cyg X-2.
The semi-coherent search method used in this paper can only create one segment per observation; where $T$ is longer than an observation, the data is zero-padded to increase resolution in the Fourier frequency domain, but does not yield increased sensitivity.
Therefore, increasing $T$ beyond that of the initial search of dataset 3 would not yield improved sensitivity, and therefore we are unable to perform a follow-up search for the candidate.
Nevertheless, we note that the Fig.~\ref{fig:dataset3_cygx2} shows features which disfavour the presence of a signal; the candidate is unresolved in the $P_{\text{orb}}$ parameter space (compare the lower left panel to that of Fig.~\ref{fig:simulatedresults}) and similar, only slightly smaller peaks appear at different values of $a$ and $T_{\text{asc}}$ (right-hand panels).
Finally, the analysis of a neighbouring segment of data (dataset 2, see Fig.~\ref{fig:cygx2_dataset2}) did not yield a strong candidate at these parameters.

The results of searches of the other Sco~X-1 and Cyg~X-2 datasets listed in Tables~\ref{tab:scox1results} and~\ref{tab:cygx2results} are given in Appendix~\ref{sec:additional-searches}.

\subsection{Assuming no spin wandering}

We also performed searches for each source assuming no spin wandering. 
The most sensitive searches were limited by computational cost and achieved fractional pulse amplitude upper limits of 0.034\% for Sco~X-1 (Fig.~\ref{fig:scox1_no_spin}) and  0.23\% for Cyg~X-2 (Fig.~\ref{fig:cygx2_no_spin}). 
These searches gave the most stringent upper limits on the detectable fractional pulse amplitude for the respective sources.

\section{Recovery of a simulated signal}\label{SS:simulatedresults}

\begin{table}
\caption{Simulated and recovered best-match parameters for the simulated search.}
\begin{tabularx}{\linewidth}{Xccc}
    \hline
    Parameter & Simulated & Recovered & Units \\
    \hline
     $\nu$     & 502.4000 $\pm$  0.0001 (95\%)  &  502.40  & Hz \\
     $P_\text{orb}$   & [68023, 68024]    & 68023.4     & s \\
     $a$         & 2.750 $\pm$   0.003 (95\%) &  2.744   & lt-s \\
     $T_\text{asc}$ & 600081300 $\pm$ 20 (95\%) & 600081290 & GPS s \\
     $A$   & 693.21 $\pm$ 74.47 (95\%) & 629.44 & - \\
    \hline
\end{tabularx}
\label{table:inj_params}
\end{table}

\begin{figure}
  \includegraphics[width=\columnwidth]{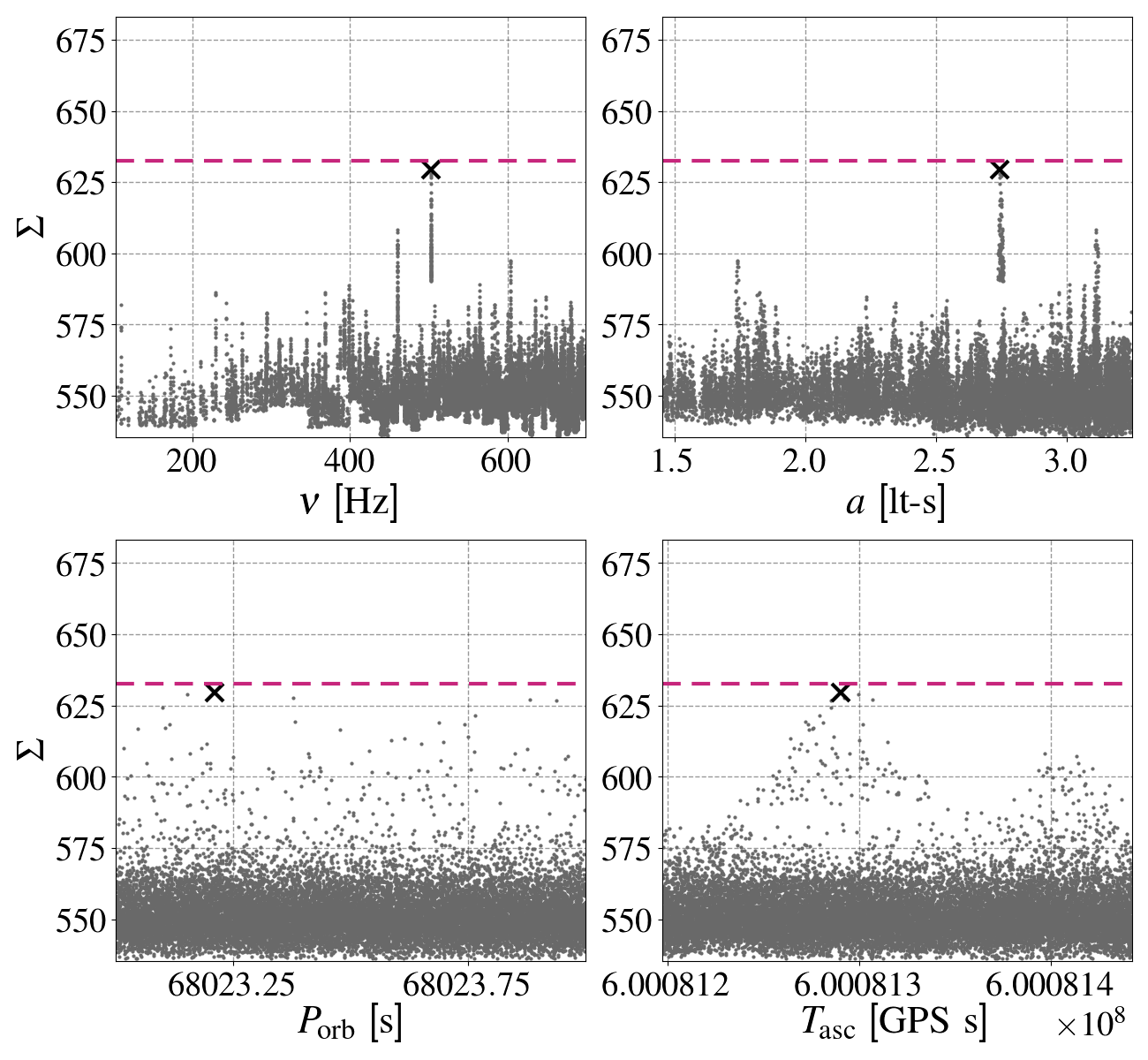}
  \caption{Detection statistic $\Sigma$ plotted against the search range of frequency $\nu$, projected semi-major axis $a$, period $P_{\text{obs}}$ and time of ascension $T_{\text{asc}}$ for a simulated dataset, with similar properties to the Sco~X-1 dataset with no spin wandering, and containing a signal with a fractional pulse amplitude of 0.034\%. The black cross indicates the template with the largest detection statistic found by the search. The dashed horizontal line at $\Sigma ^{1\%} = 632.59$ indicates the 1\% false alarm threshold on the detection statistic and corresponds to the same fractional pulse amplitude as the simulated signal. The 100 largest detection statistics recorded in each parameter space partition are plotted.}
  \label{fig:simulatedresults}
\end{figure}

To illustrate the recovery of X-ray pulsations at the very limit of search sensitivity, we performed a search of an simulated dataset.
The dataset has similar properties to the Sco~X-1 dataset with no spin wandering, including the time-span, duty cycle, background photon count rate.
It also contains a simulated signal, generated using the signal model described by Eq.~\ref{eq:signal}, with a fractional pulse amplitude of 0.034\%
The frequency and orbital parameters of the simulated signal are randomly chosen from the Sco~X-1 parameter ranges given in Table~\ref{tab:parameter_scox1}. 

The simulated signal is just recovered at the 1\% false alarm threshold, as shown in Fig.~\ref{fig:simulatedresults}, demonstrating that the method can detect signals with the claimed fractional pulse amplitude of 0.034\% at 90\% confidence.
The frequency and projected semi-major axis parameters of the recovered signal are clearly constrained. The uncertainties in orbital period and time of ascension from observations are small enough, however, that the search does not provide tighter constraints.
The parameters of the simulated signal and of the recovered best-match template are given in Table~\ref{table:inj_params}. 
The recovered signal is not identical to the simulated signal; this discrepancy is expected since there is a loss of signal-to-noise ratio due to the mismatch between the signal and (random) template parameters. 

\section{Discussion}\label{sec:discussion}

We conducted a semi-coherent search for X-ray pulsations on archival {\it RXTE}/PCA data for Sco~X-1 and Cyg~X-2 with a total on-source observation time of $\sim \SI{104.4}{hr}$ and $\sim \SI{48.5}{hr}$ respectively.
We perform a set of searches assuming that the spin wandering timescale limits the length of the coherently-analysed segments, and an additional set assuming no restrictions due to spin wandering.
We detected no clear evidence of persistent X-ray pulsations, and found the most stringent upper limits (at 90\% confidence) on the fractional pulse amplitude to be 0.034\% for Sco~X-1 and 0.23\% for Cyg~X-2.

\cite{Vaughan1994} performed fully-coherent searches of 1024--$\SI{2048}{s}$ observations taken by the Large Area Counter onboard the {\it Ginga} satellite, and set best upper limits (at 99\% confidence) of 0.28\% on Sco~X-1 and 0.37\% on Cyg~X-2.
The time-spans of the coherent searches in~\cite{Vaughan1994} are comparable to the coherent segment lengths of our searches for Cyg~X-2 (Table~\ref{tab:cygx2results}), and up to a factor of $\sim 3.5$ greater than those of our searches for Sco~X-1 (Table~\ref{tab:scox1results}).
On the other hand, our semi-coherent searches analyse $\sim 29$ times more than data in total than a single coherent analysis of~\cite{Vaughan1994}, and moreover combine coherent segments together for improved sensitivity.
Taken together with the improved sensitivity of {\it RXTE} compared to {\it Ginga} -- $\sim 0.17$ mCrab~\citep{Jahoda2006} versus $\sim 0.2$ mCrab~\citep{Turner1989} -- our upper limits on Sco~X-1 and Cyg~X-2 improve upon those of~\cite{Vaughan1994} by factors of $\sim 8.2$ and $\sim 1.6$ respectively.

The non-detection of X-ray pulsations from Sco~X-1 and Cyg~X-2, presented in this paper, is consistent with null results from other pulsation searches for LMXBs~\citep[e.g.][]{Messenger_Patruno2015,Patruno2018}.
The absence of observed pulsations from many LMXBs might be due to a number of factors:
\begin{enumerate}
    \item Weak magnetic fields that do not channel matter would present a possible explanation for the absence of pulsations~\citep{Cumming2001}. Some LMXBs such as Aquila X-1, however, exhibit stong transient pulsations~\citep{Casella2008, Messenger_Patruno2015}, which would require a momentary increase in the magnetic field strength inconsistent with this model.
    \item Pulsations may be present but not persistent. Some neutron stars are known to intermittently exhibit X-ray pulsations~\citep{Galloway2007}. The semi-coherent search method used in this paper is designed for persistent signals, and would be unlikely to detect pulsations that are both weak and intermittent.
    \item Pulsations may be suppressed by electron scattering in the optically thick material surrounding the LMXB~\citep{Titarchuk2002,Gogus2007}, and therefore may be too weak to be detectable with current data analysis techniques, or else absent altogether.
    \item Sco~X-1 may not be a neutron star, in which case we would not expect X-ray pulsations.
\end{enumerate}

Future searches for X-ray pulsations from Sco~X-1, Cyg~X-2, and other LMXBs could make use of improved data analysis methods that e.g.\ permit longer segments of data to be analysed coherently.
In addition, future X-ray satellite missions such as STROBE-X~\citep{STROBE-X} may provide more sensitive X-ray data to search for X-ray pulsations and thereby support the search for continuous gravitational waves.

\section*{Acknowledgements}

This research was conducted by the Australian Research Council Centre of Excellence for Gravitational Wave Discovery (OzGrav), through project number CE170100004.
The authors also gratefully acknowledge the Science and Technology Facilities Council of the United Kingdom. CM is supported by the Science and Technology Research Council (grant No. ST/L000946/1).
This research has made use of data and software provided by the High Energy Astrophysics Science Archive Research Center (HEASARC), which is a service of the Astrophysics Science Division at NASA/GSFC and the High Energy Astrophysics Division of the Smithsonian Astrophysical Observatory.
The searches for X-ray pulsations were performed on the Oz\-STAR national facility at Swinburne University of Technology. Oz\-STAR is funded by Swinburne University of Technology and the National Collaborative Research Infrastructure Strategy (NCRIS). 

\section*{Data availability}

The data used in this paper was processed from data provided by the High Energy Astrophysics Science Archive Research Center (HEASARC).
Links to processed data and result files can be found at: \href{https://github.com/shanikagalaudage/xray_pulsations}{https://github.com/shanikagalaudage/xray\_pulsations}

\bibliographystyle{mnras}
\bibliography{refs}

\appendix

\section{Dataset observation IDs}\label{sec:dataset-IDs}

\begin{table}
\caption{Observation IDs for Scorpius X-1}
\begin{tabularx}{\linewidth}{lp{0.5\linewidth}r}
    \hline
    Dataset & Observation IDs & Duty Cycle (\%)\\
    \hline
    1 & 10061-01-03-00, 20053-01-02-00, 20053-01-02-01, 20053-01-02-02, 20053-01-02-03, 30036-01-01-00, 20053-01-02-04, 20053-01-02-05, 30036-01-02-00 & 16.34\\
    \hline
    2 & 30035-01-01-00, 30035-01-02-00, 30035-01-05-00, 30035-01-03-00, 30035-01-06-00, 30035-01-04-00 & 16.15 \\
    \hline
    3 & 30035-01-07-00, 30035-01-08-00, 30035-01-11-00, 30035-01-09-00, 30035-01-10-00. & 11.09\\
    \hline
    4 & 40706-02-01-00, 40706-02-03-00, 40706-02-06-00, 40706-01-01-00, 40706-02-09-00, 40706-02-08-00, 40706-02-10-00, 40706-02-12-00, 40706-02-13-00, 40706-02-14-00, 40706-01-02-00, 40706-02-16-00, 40706-02-17-00, 40706-02-18-00, 40706-02-19-00, 40706-02-21-00, 40706-02-22-00, 40706-02-23-00, 40706-01-03-00 & 22.84 \\
    \hline
    no spin & 40020-01-02-00, 40020-01-01-00, 40020-01-01-01, 40020-01-01-02, 40020-01-01-03, 40020-01-01-04, 40020-01-01-05, 40020-01-01-06, 40020-01-01-07, 40020-01-03-00, 40020-01-03-01 & 16.19 \\
    \hline
\end{tabularx}
\label{tab:obs_ids_scox1}
\end{table}

\begin{table}
\caption{Observation IDs for Cygnus X-2}
\begin{tabularx}{\linewidth}{lp{0.5\linewidth}r}
    \hline
    Dataset & Observation IDs & Duty Cycle (\%)\\
    \hline
    1 & 20053-04-01-00, 20053-04-01-01, 20053-04-01-02, 20053-04-01-06, 20053-04-01-03, 20053-04-01-04, 20053-04-01-07, 20053-04-01-05 &  25.98\\
    \hline
    2 & 95345-01-01-00, 95345-01-02-00, 95345-01-02-01, 95345-01-03-00, 95345-01-04-00, 95345-01-05-00, 95345-01-06-01, 95345-01-06-00, 95345-01-07-00, 95345-01-08-00, 95345-01-08-01, 95345-01-09-00, 95345-01-10-00, 95345-01-11-00, 95345-01-12-00, 95345-01-13-00, 95345-01-14-00, 95345-01-15-00, 95345-01-16-00, 95345-01-17-00, 95345-01-18-00, 95345-01-19-00, 95345-01-20-00, 95345-01-21-00 & 13.75\\
    \hline
    3 & 95345-01-22-00, 95345-01-23-00, 95345-01-23-01, 95345-01-24-00, 95345-01-25-00, 95345-01-26-00, 95345-01-26-01, 95345-01-27-00, 95345-01-28-00, 95345-01-29-00, 95345-01-29-01, 95345-01-30-00, 95345-01-31-00, 95345-01-32-00, 95345-01-32-02, 95345-01-33-00, 95345-01-34-00, 95345-01-34-01, 95345-01-35-02, 95345-01-35-01, 95345-01-36-00, 95345-01-37-00, 95345-01-37-01, 95345-01-38-00, 95345-01-38-01, 95345-01-39-00, 95345-01-40-01 & 12.3\\
    \hline
    no spin & Datasets 2 and 3 & 4.71\\
    \hline
\end{tabularx}
\label{tab:obs_ids_cygx2}
\end{table}

Tables~\ref{tab:obs_ids_scox1} and~\ref{tab:obs_ids_cygx2} list the observation IDs for the Sco~X-1 and Cyg~X-2 datasets, respectively, that were analysed in this paper. The duty cycle is the measured as the percentage exposure time over the total time span of the observation set.

\section{Additional Sco X-1 and Cyg X-2 searches}\label{sec:additional-searches}

\begin{figure}
  \includegraphics[width=\columnwidth]{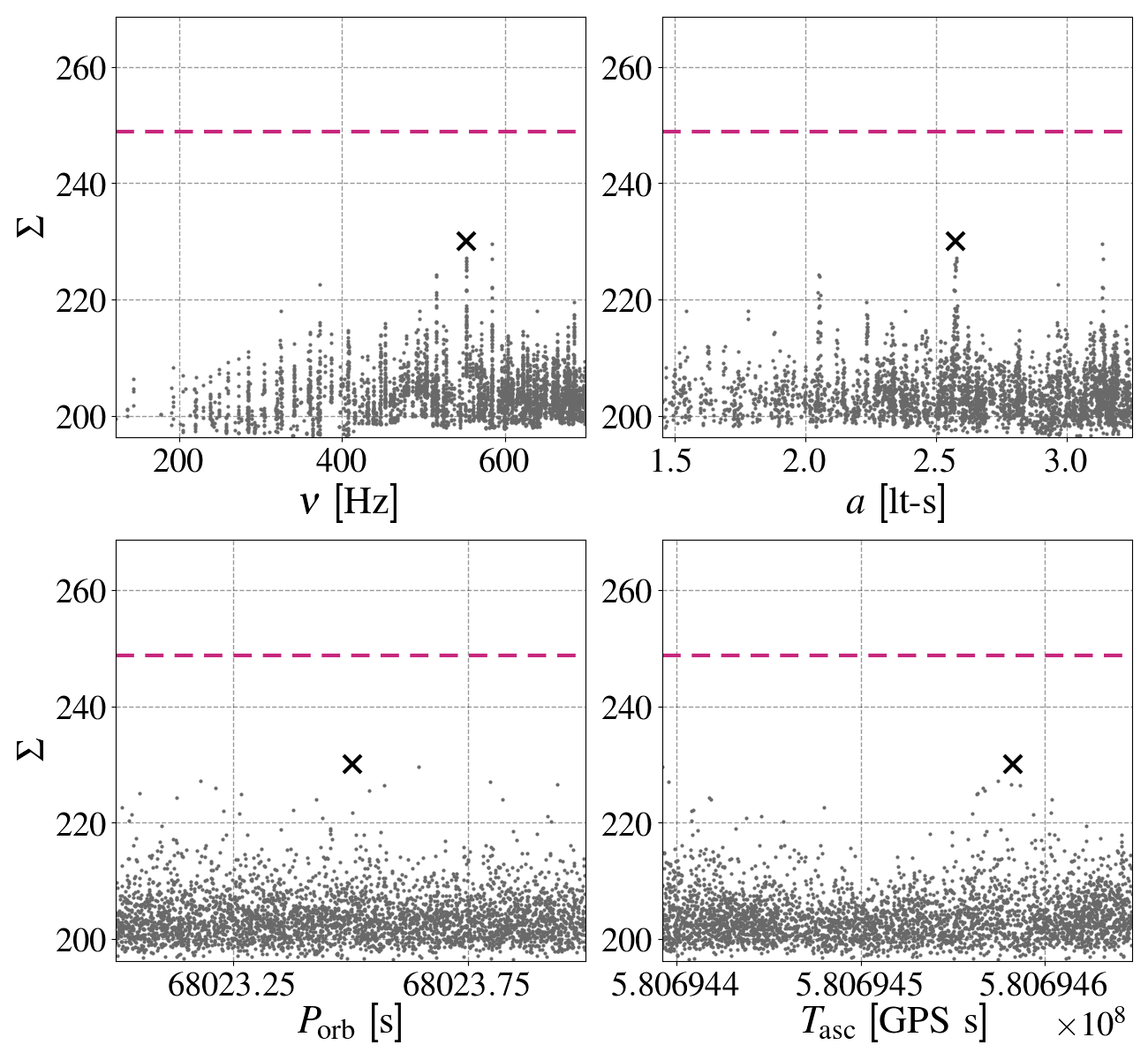}
  \caption{Results for Sco~X-1 dataset 2: $\Sigma ^{1\%} = 248.78$.}
  \label{fig:scox1_dataset1}
\end{figure}

\begin{figure}
  \includegraphics[width=\columnwidth]{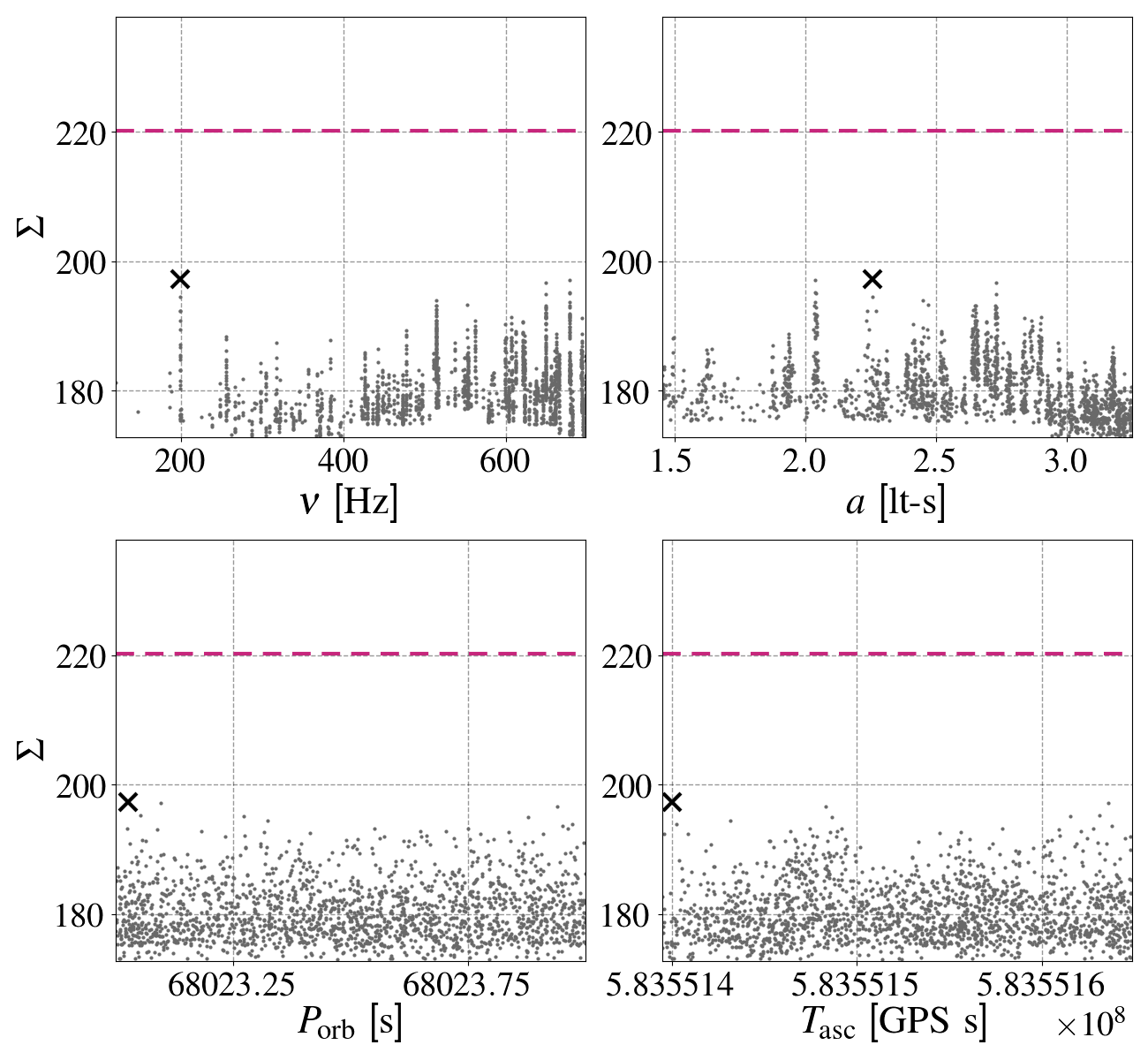}
  \caption{Results for Sco~X-1 dataset 3: $\Sigma ^{1\%} = 220.22$.}
  \label{fig:scox1_dataset3}
\end{figure}

\begin{figure}
  \includegraphics[width=\columnwidth]{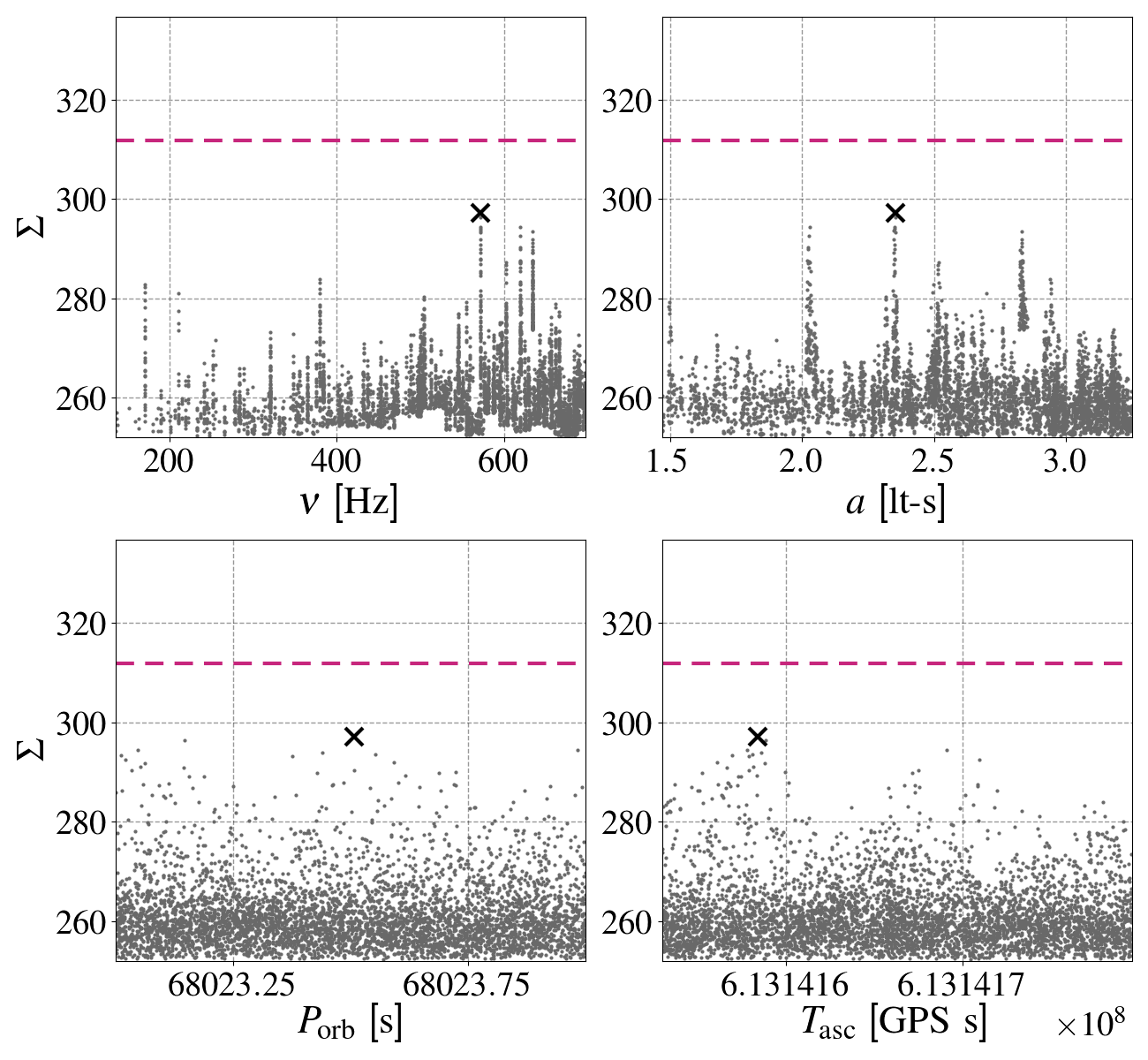}
  \caption{Results for Sco~X-1 dataset 4: $\Sigma ^{1\%} = 311.79$.}
  \label{fig:scox1_dataset4}
\end{figure}

\begin{figure}
  \includegraphics[width=\columnwidth]{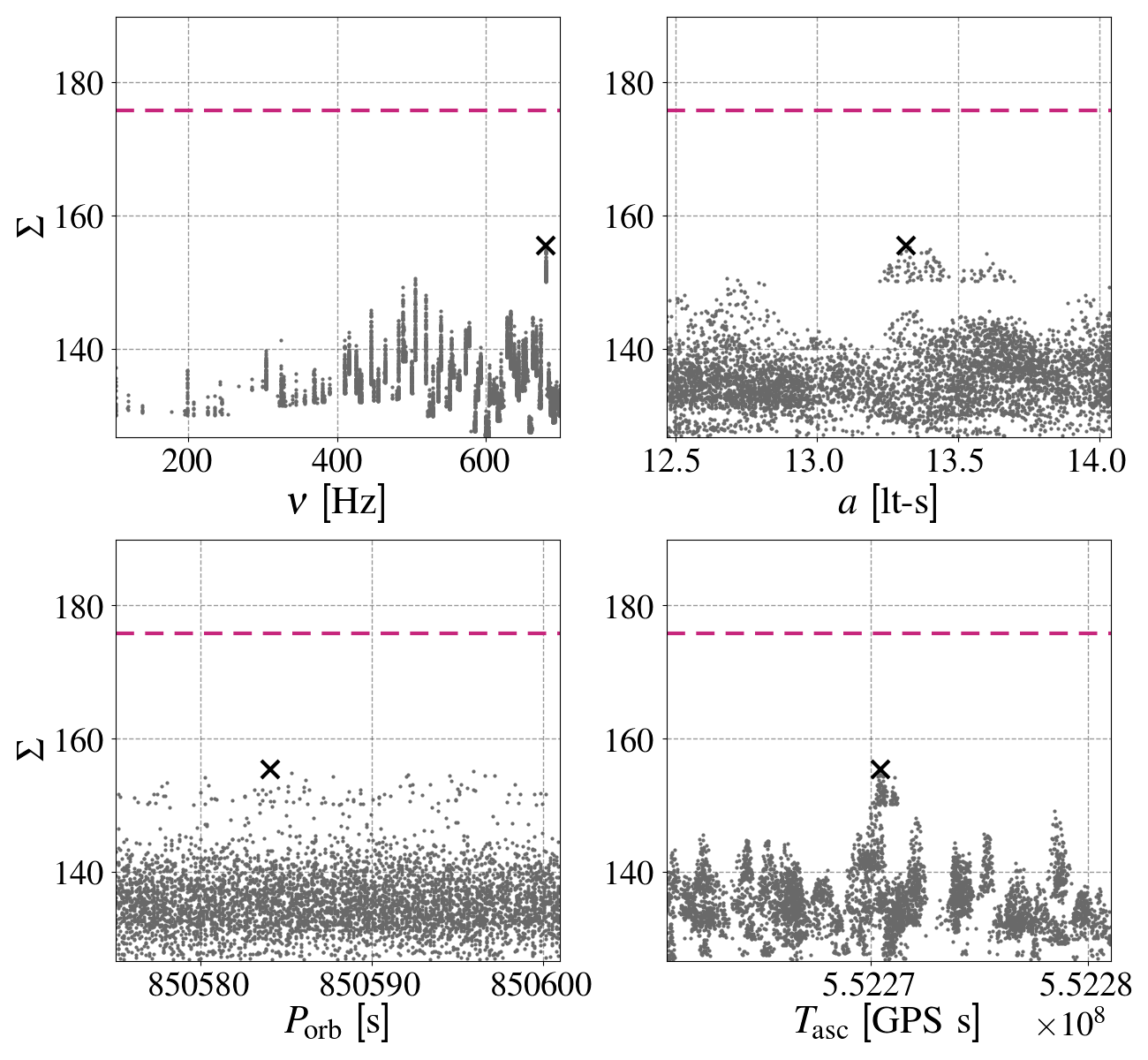}
  \caption{Results for Cyg~X-2 dataset 1: $\Sigma ^{1\%} = 175.80$.
  }
  \label{fig:cygx2_dataset1}
\end{figure}

\begin{figure}
  \includegraphics[width=\columnwidth]{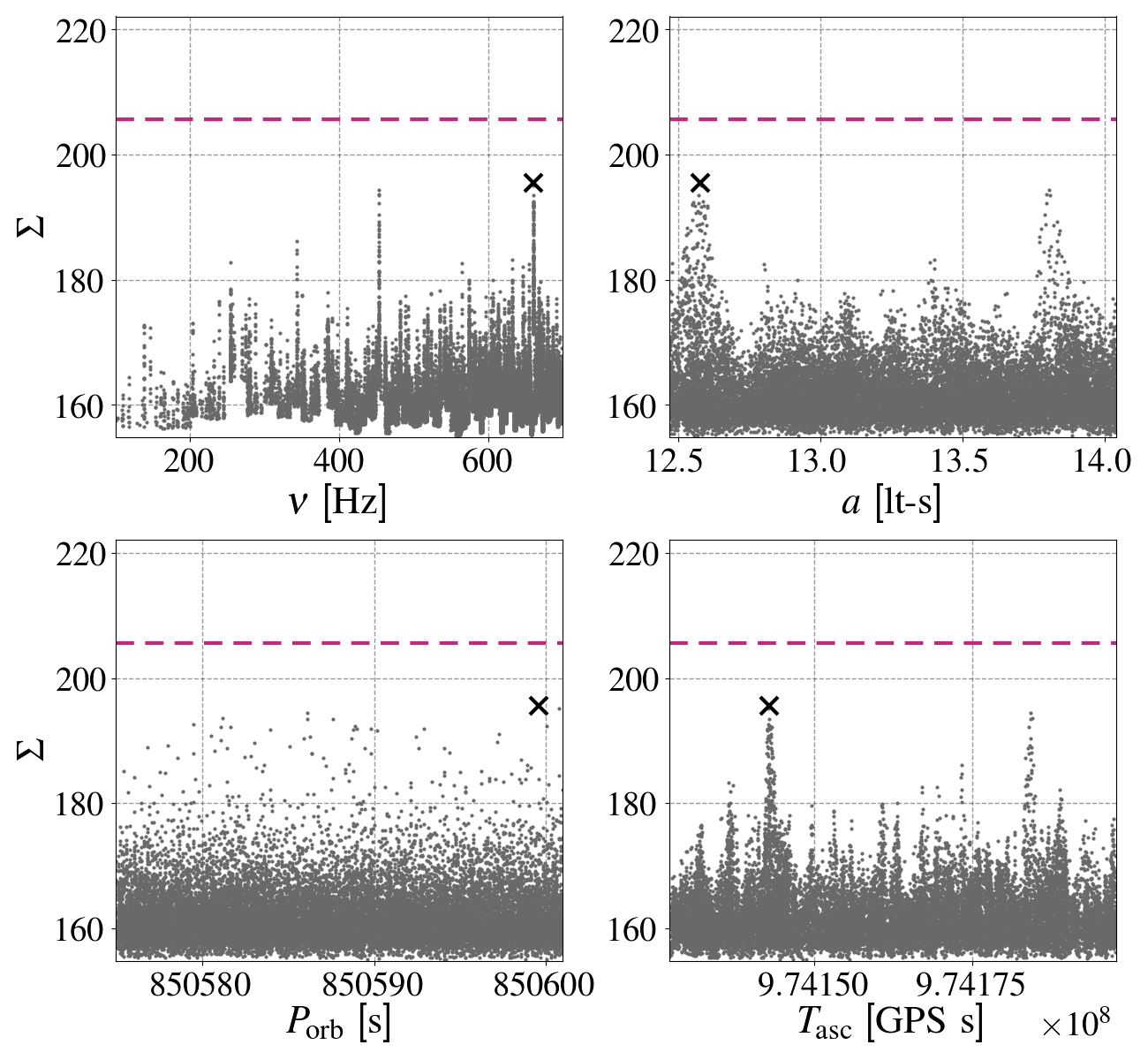}
  \caption{Results for Cyg~X-2 dataset 2: $\Sigma ^{1\%} = 279.57$.
  }
  \label{fig:cygx2_dataset2}
\end{figure}

In this section we present the results from the additional datasets analysed. For each figure we plot the detection statistic $\Sigma$ against the search range of frequency $\nu$, projected semi-major axis $a$, period $P_{\text{obs}}$ and time of ascension $T_{\text{asc}}$. The black cross indicates the template with the largest detection statistic found by the search ($\Sigma ^{*}$). The dashed horizontal line indicates the 1\% false alarm threshold on the detection statistic ($\Sigma ^{1\%}$). The 100 largest detection statistics recorded in each parameter space partition are plotted.

\bsp	
\label{lastpage}
\end{document}